\documentstyle[12pt,aaspp4]{article}

\def\icm{${\rm cm}^{-1}$}
\def\he3{$^3{\rm He}$}

\def\dminus{{\it92$-$94}}
\def\dplus{{\it92$+$94}}

\def\kelvin#1{$#1 \,{\rm K}$}
\def\db#1{$#1 \,{\rm dB}$}

\begin{document}

\slugcomment{Submitted to {\em Ap. J. Letters}; \quad {\tt astro-ph/9603017} }

\title{A CMBR Measurement Reproduced: A Statistical Comparison of
MSAM1-94 to MSAM1-92}

\author{
C.~A.~Inman\altaffilmark{1},
E.~S.~Cheng\altaffilmark{2},
D.~A.~Cottingham\altaffilmark{3},
D.~J.~Fixsen\altaffilmark{4},
M.~S.~Kowitt\altaffilmark{2},
S.~S.~Meyer\altaffilmark{1},
L.~A.~Page\altaffilmark{5},
J.~L.~Puchalla\altaffilmark{2},
J.~E.~Ruhl\altaffilmark{6},
and~R.~F.~Silverberg\altaffilmark{2}}

\altaffiltext{1}{University of Chicago, 5640 S. Ellis St., Chicago, IL
60637}
\altaffiltext{2}{Laboratory for Astronomy and Solar Physics, 
NASA/Goddard Space Flight Center, Code 685.0, Greenbelt, MD 20771}
\altaffiltext{3}{Global Science and Technology, Inc., NASA/GSFC Laboratory
for Astronomy and Solar Physics, Code 685.0, Greenbelt, MD 20771}
\altaffiltext{4}{Applied Research Corporation, NASA/GSFC Laboratory for
Astronomy and Solar Physics, Code 685.3, Greenbelt, MD 20771}
\altaffiltext{5}{Princeton University Physics Dept., Princeton, NJ
08544}
\altaffiltext{6}{Department of Physics, University of California,
Santa Barbara, CA 93106}

\begin{abstract}

The goal of the second flight of the Medium Scale Anisotropy
Measurement (MSAM1-94) was to confirm the measurement of cosmic
microwave background radiation (CMBR) anisotropy made in the first
flight (MSAM1-92).  
The CMBR anisotropy and interstellar dust emission signals
from the two flights are compared by forming the sum and
difference of those portions of the data with the same pointings on the
sky.  The difference data are consistent with a null detection, while
the summed data show significant signal.  We conclude that MSAM1-92
and MSAM1-94 measured the same celestial signal.
 
\end{abstract}

\section{Introduction}

Measurements of anisotropy in the Cosmic Microwave Background
Radiation (CMBR) continue as a subject of considerable interest to the
astrophysics community.  Future anisotropy measurements on scales of
0\fdg1 to 1\fdg0 will discriminate among early universe models and
determine fundamental cosmological parameters (e.g. \cite{hu96a},
\cite{knox95b} and \cite{jungman95}).

Measurements of anisotropy at angular scales near 0\fdg5 have been
reported recently by \cite{ruhl95}, \cite{netterfield96},
\cite{gundersen94}, and \cite{tanaka95}.  \cite{wilkinson95} voiced a
common concern when he pointed out that ``there are plausible
systematic effects at levels comparable with the reported
detections.''  
To address this concern the 1994 flight of the Medium
Scale Anisotropy Measurement (MSAM1) observed the same field as the
1992 flight to limit the possibility of systematic sources
of the signal.

\cite{cheng94} (hereafter Paper~I) reported observations of anisotropy
in the CMBR from the first flight of MSAM1 in 1992 (MSAM1-92).  
\cite{cheng95} (hereafter Paper~II) reported the results from 
the second flight in 1994 (MSAM1-94).  
A conclusion of the latter is that while a quantitative
comparison was pending, there was good qualitative agreement between
the two flights in the double difference data set, and that agreement
was inconclusive for the single difference data set.  This Letter
presents a quantitative comparison of the MSAM1-92 and MSAM1-94 data
sets.

\section{Instrument and Observations}

The MSAM1 instrument has been fully described in \cite{fixsen96a}
(hereafter Paper~III); only an overview is given here.  It is an
off-axis Cassegrain telescope with a 4-channel bolometric radiometer
at the focus.  The beamsize is 28\arcmin\ FWHM and is moved
$\pm40\arcmin$ on the sky by the nutating secondary.  The radiometer
has 4 frequency channels placed at 5.7, 9.3, 16.5, and 22.6 \icm.  For
these observation, emission in the lower two channels is dominated by
the \kelvin{2.7} CMBR, while $\sim$ \kelvin{20} interstellar dust
dominates the two higher channels.

The instrument configuration was similar for the two flights,
with changes made only to
the warm signal electronics and the gondola structure.
These changes are discussed extensively in Paper~III; the
modifications to the electronics improved the noise performance, while
those to the gondola reduced sidelobe sensitivity.  The original
superstructure had a large reflecting area above the beam, from which
earthshine could potentially diffract into the beam.  For the second
flight, the gondola was suspended by a cable system which reduced the
far-sidelobe response.  The measured near sidelobe response dropped
from \db{-55} in 1992 in the worst case to less than \db{-75} in all
cases in 1994.

As described in Papers~I and II, the observed field is two strips at
declination $81\fdg8 \pm 0\fdg1$, from right ascension 15\fh27 to
16\fh84, and from 17\fh57 to 19\fh71 (all coordinates are J1994.5).
Fig.~\ref{f_fields} shows the weighted beam centers of the fields
observed in the 1992 and 1994 flights.  A CCD camera is used to
determine absolute pointing for both flights.  The final accuracy of
the pointing determination is 2\farcm5, limited by the gyroscope
signal interpolation.  This 2\farcm5 uncertainty is small compared to
the size of our beam (28\arcmin) and the bins (14\arcmin) used below,
ensuring adequate alignment of the two datasets.

During both flights Jupiter was observed to calibrate the instrument
and map the telescope beam.  Beam maps and calibrations are done
separately for the two flights.  The shape of the beam map is
determined to 4\% of the maximum amplitude.  Random noise in the
gyroscope system contributes 3.5\%, and cosmic rays striking the detectors
contribute 1.5\%.  Also, the choice of smoothing algorithm causes a
1.5\% systematic effect.  
Combining this 4\% error from each flight gives a
5.8\% relative calibration uncertainty.  
The uncertainty in Jupiter's
intrinsic brightness leads to an additional systematic uncertainty of
10\% for the results of each flight; however, except for possible time
variations in Jupiter's brightness which we do not expect, this
uncertainty does not contribute to the comparison of the two flights
discussed here.

\section{Reanalysis of 1992 Data}

The analyzed data sets reported in Papers ~I and II cannot be directly
compared for two reasons: 1) the boundaries of the sky bins are
different, and 2) the analysis reported in Paper~I neglects
correlations introduced by the removal of the small offset drift.  The
1992 data is reanalyzed to account for these correlations, using a
procedure nearly identical to that of Paper~II.  The 1994 data is also
reanalyzed, with differences from the original analysis noted in the
text below.  Here we first review the Paper~II analysis, then note the
differences between that and the reanalysis used here.

First, the cosmic ray events are removed from the time stream.  Cosmic
ray removal techniques are different for the two years, and are
discussed in Papers~I and II.  The data are then analyzed in a manner
that provides sensitivity to two different angular scales on the sky.
This is done by weighting the the time stream, $S_i$, with one of two
demodulation templates, $d_i$, giving one ``demodulated data point'',
$\Delta T_{cycle} = \sum_{i} d_i S_i$, for each full cycle of the
secondary mirror movement.  The ``single difference demodulation''
weights the secondary-left data positively while weighting the
secondary-right data negatively, giving a $\Delta T$ equal to the
difference between the left and right temperatures. The result is a
two lobed beam pattern on the sky with 80\arcmin\ beam separation.
The ``double difference demodulation'' weights the secondary-centered
data positively, while weighting the secondary-left and
secondary-right data negatively, giving a $\Delta T$ equal to the
difference between the center and the side temperatures. This produces
a three lobed beam pattern on the sky, with 40\arcmin\ beam
separation.  The single difference and double difference data are
nearly statistically independent.

A linear model is fit to these demodulated data including intensity
for each sky bin and slow drifts in time.  The noise used in the fit
is estimated from the data.  Both the time drift and noise estimate
are described further below.  The results of the linear fit are signal
amplitudes for each sky bin with their associated covariances.

A spectral model for each sky bin consisting of CMBR anisotropy plus
emission from \kelvin{20} Galactic dust, with emissivity proportional to
frequency to the 1.5 power, is fit to all four frequency channels of
binned sky data.  The results of this spectral fit are the intensity
of a ``DUST'' component and a ``CMBR'' component in each sky bin.  The
differences between the Paper~II analysis and that done for this paper
follow.

\begin{figure}[tbhp]
\plotone{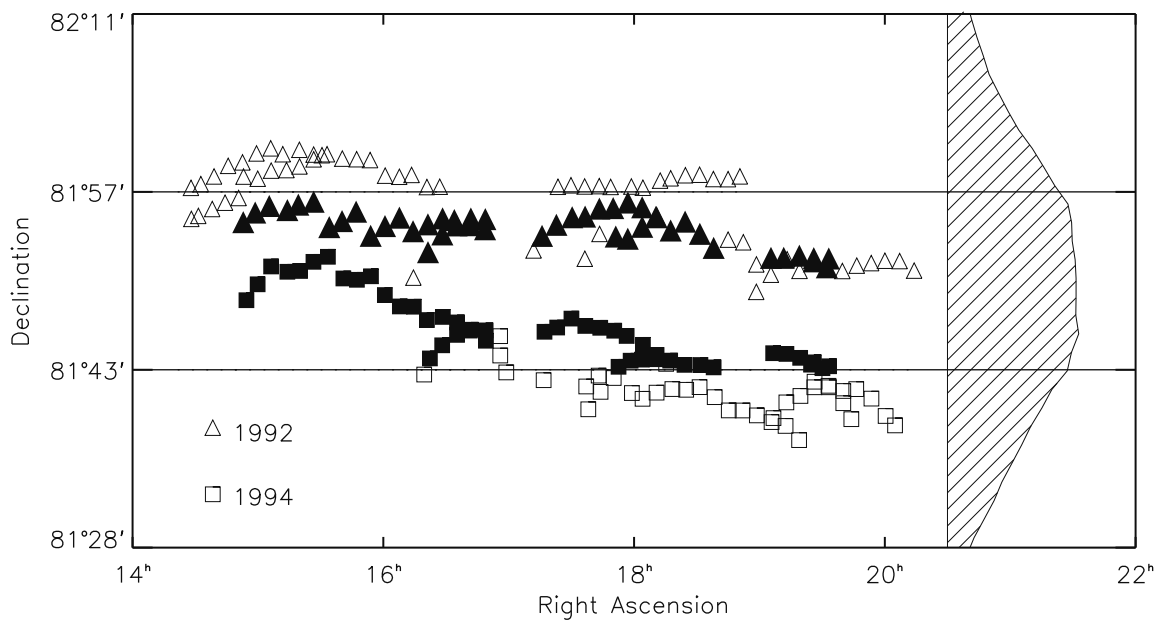} \\[2ex]
\caption{The weighted positions for each sky bin for both years.
The triangles mark the 1992 centers and the squares
mark the 1994 centers.  The declination scale has been greatly
expanded relative to the RA scale in order to see the detailed
pointing differences.  The filled symbols are the weighted centers for
the bins used in the comparison ($\langle \delta \rangle = $ 81\arcdeg
50\arcmin). The bin boundaries (every 0\fdg24, or 0\fh11) in RA are not 
shown.  The declination bin boundaries (every 0\fdg24) are marked by the
horizontal lines.  The angular orientation is ignored in this plot,
but is not in our analysis.  The vertical beam profile is plotted in
the hatched region. Note that at this declination, every hour of RA
corresponds to about 2\arcdeg. }
\label{f_fields}
\end{figure}

This analysis uses a 0\fdg24 bin size, double the size used in
Paper~II, which corresponds to the size of the central beam plateau.
Angular orientation bins, which account for sky rotation relative to
the secondary chop axis, are 20\arcdeg, also double the previous size.
The weighted beam centers of the identical bins are shown as filled
symbols in Fig.~\ref{f_fields}.

The noise estimates are formed from demodulated data.  This is a
change from Paper~II, where the estimate is made after after having
removed the drift model.  The noise estimates are made separately for
each minute of data by measuring the rms of the demodulated data in
that minute.  The new noise estimate is used in reanalyzing both the
1992 and 1994 datasets.  
True sky signals make a negligible contribution to this rms estimate over
these short time scales.
This change has no substantial effect on the
results of this Letter.

Also, in Paper~II the drift model included terms based on gondola
sensors (air pressure, and the pitch and roll angles of the gondola
outer frame).  This model was used in the 1994 reanalysis, while it
was not used for the 1992 reanalysis.  Instead, the original model for
the drifts described in Paper~I, a spline with knots every 2.5
minutes, was used.

\section{Comparison}

We compare the signal measured at each point on the sky as measured in
the two flights, not just the rms levels of the sky signal found in
each data set.  In the 1994 flight, we attempted to observe the
identical swath of sky observed in the 1992 flight.  As can be seen in
Fig.~\ref{f_fields}, which is extremely enlarged in declination
relative to right ascension, the 1994 flight was low by about
10\arcmin\ .  To enable direct comparison, only the data from those
bins which fall into the center declination bin is used.  After this
selection $\sim$50\% of the data is retained.  The data from the 1992
flight is differenced from that of the 1994 flight to form a
difference data set, \dminus\ .  Similarly, the two data sets are
summed to form a sum data set, \dplus\ .  This is done for each
demodulation and for both CMBR and DUST.  To allow for differing
offsets in the two flights, a weighted mean is removed from each
dataset.  The covariance matrix, $V_{ij}$, for both the sum and
difference sets is the sum of the masked 1992 and 1994 covariance
matrices.  There is no cross term because the flights have independent
noise.  The significance of any detected signal in the sum or
difference is tested with a $\chi^2$ statistic,
\[
\chi^2_{\pm} = \sum_{ij}
({\it92 \pm 94})_i V^{-1}_{ij} ({\it92 \pm 94})_j .
\]

The $\chi^2$ and degrees of freedom, and the cumulative probability,
$P(\chi^2)$, for the comparison is shown in Table~\ref{t_chi}.  $P$ is
the probability of getting a value of $\chi^2$ at or above the
observed value, under the assumption that there is no signal in the
data.  This should be the case for the difference data, where the
common sky signal should cancel.

To check the effect of the relative calibration uncertainty on
$\chi^2$, the 1994 dataset is rescaled by $\pm 6\%$ and the value of
$\chi^2$ recalculated.  In all cases $|\Delta \chi^2| \leq 2$.

A Kolmogorov-Smirnov (KS) test (\cite{press92}) of the \dminus\
probabilities (.04, .22, .41, and .91) gives a 74\% probability that
these are drawn from a uniform distribution from 0 to 1.  Based on
this, we conclude that the \dminus\ data in both the single and double
difference demodulations for both the CMBR and DUST components is
consistent with no observed signal.

A KS test of the \dplus\ probabilities ($2\times 10^{-12}$, $2\times
10^{-8}$, $4\times 10^{-7}$, and $1\times 10^{-4}$) gives a $7\times
10^{-4}$ probability that these are drawn from a uniform distribution
from 0 to 1.  From this, together with the extremely low $\chi^2$
probabilities themselves, we see that there are statistically
significant signals in all four \dplus\ datasets.  This result,
combined with the absence of such signals in the \dminus\ datasets,
enables us to conclude that the signals observed during the two
flights are common, and therefore present on the sky.

\begin{deluxetable}{rlcc}
\tablecolumns{4}
\tablecaption{Comparison of 1992 and 1994 Data Sets \label{t_chi} }
\tablehead{
\colhead{Type} &
    \colhead{Data Set} & \colhead{$\chi^2$/{\small DOF}} & \colhead{$P$}}

\startdata
\cutinhead{Single Difference}
        CMBR &   \dminus  &   52 / 45 & 0.22 \nl
             &   \dplus  &   89 / 45 & $1\times 10^{-4}$\nl
\nl
        DUST &   \dminus  &   47 / 45 & 0.41 \nl
             &   \dplus  &  145 / 45 & $2\times 10^{-12}$\nl
\cutinhead{Double Difference}
        CMBR &   \dminus  &   33 / 45 & 0.91 \nl
             &   \dplus  &  118 / 45 & $2\times 10^{-8}$\nl
\nl
        DUST &   \dminus  &   63 / 45 & 0.04 \nl
             &   \dplus  &  109 / 45 & $4\times 10^{-7}$\nl
\enddata
\end{deluxetable}

\section{Conclusions}

The same region of the sky was observed in the 1992 and 1994 flights
to confirm the detection of a celestial signal.  It is clear from the
statistical analysis that the same sky signal is measured in these two
flights.  We conclude that at the level of our signal, our
measurements are 
likely to be 
free from significant contamination from time-varying
systematics such as sidelobe pickup or atmospheric contamination.

In addition to our own confirmation of the MSAM1-92 results, the
Saskatoon experiment has recently observed this section of sky at
lower frequencies, 36~GHz to 46~GHz (\cite{netterfield96}).  They have
compared their signal with the double difference CMBR signal from
Paper I, and find good agreement.  This result, spanning nearly a
decade in frequency, is strong evidence that we are observing CMBR
anisotropies rather than some other astrophysical foreground source.

\acknowledgments

We would like to thank E. Magnier, R. Rutledge, L. Knox, and A. Goldin
for useful conversations.  The research was supported by the NASA
Office of Space Science, Astrophysics Division through grants NTG
50720 and 50908 and RTOP 188-44.

\clearpage


\clearpage

\begin{thebibliography}{}

\bibitem[Cheng {\em et~al.} 1994]{cheng94}
Cheng, E.~S. {\em et~al.} 1994, \apjl, {\bf 422}, L37.

\bibitem[Cheng {\em et~al.} 1996]{cheng95}
Cheng, E.~S. {\em et~al.} 1996, \apjl, {\bf 456}, L71.

\bibitem[Fixsen {\em et~al.} 1996]{fixsen96a}
Fixsen, D.~J. {\em et~al.} 1996, \apj.
\newblock submitted, preprint {\tt astro-ph/9512006}.

\bibitem[Gundersen {\em et~al.} 1995]{gundersen94}
Gundersen, J.~O. {\em et~al.} 1995, \apjl, {\bf 443}, L57.

\bibitem[Hu and White 1996]{hu96a}
Hu, W. and White, M. 1996, \apj.
\newblock submitted, preprint {\tt astro-ph/9602019}.

\bibitem[Jungman {\em et~al.} 1995]{jungman95}
Jungman, G., Kamionkowski, M., Kosowsky, A., and Spergel, D. 1995, \prd.
\newblock submitted, preprint {\tt astro-ph/9512139}.

\bibitem[Knox 1995]{knox95b}
Knox, L. 1995, \prd, {\bf 52}, 4307.

\bibitem[Netterfield {\em et~al.} 1996]{netterfield96}
Netterfield, C.~B., Devlin, M.~J., Jarosik, N., Page, L., and Wollack, E.~J.
  1996, \apj.
\newblock submitted, preprint {\tt astro-ph/9601197}.

\bibitem[Press {\em et~al.} 1992]{press92}
Press, W.~H., Teukolsky, S.~A., Vetterling, W.~T., and Flannery, B.~P. 1992.
\newblock {\em Numerical Recipes in {FORTRAN}: The Art of Scientific
  Computing}.
\newblock Cambridge University Press, Cambridge, 2nd edition.

\bibitem[Ruhl {\em et~al.} 1995]{ruhl95}
Ruhl, J.~E., Dragovan, M., Platt, S.~R., Kovac, J., and Novak, G. 1995, \apjl,
  {\bf 453}, L1.

\bibitem[Tanaka {\em et~al.} 1995]{tanaka95}
Tanaka, S.~T. {\em et~al.} 1995, \apjl.
\newblock submitted, preprint {\tt astro-ph/9512067}.

\bibitem[Wilkinson 1995]{wilkinson95}
Wilkinson, D. 1995.
\newblock A Warning Label for Cosmic Microwave Background Anisotropy
  Experiments.
\newblock In Astbury, A. {\em et~al.}, editors, {\em Particle Physics and
  Cosmology, Proceedings of the Ninth Lake Louise Winter Institute}, page 110,
  Singapore. World Scientific.

\end{thebibliography}
\end{document}